\documentclass[prl,aps,twocolumn,floatfix,showpacs]{revtex4}
\usepackage{graphics}
\begin{document}
\title{Environmentally-Induced Rabi Oscillations and Decoherence in Phase Qubits}
\author{Kaushik Mitra, C. J. Lobb, and C. A. R. S{\'a} de Melo}
\affiliation{Joint Quantum Institute and Department of Physics \\ 
University of Maryland College Park MD 20742}
\date{\today}

\begin{abstract}
We study decoherence effects in a dc SQUID phase qubit caused by an isolation
circuit with a resonant frequency. The coupling between the SQUID phase
qubit and its environment is modeled via the Caldeira-Leggett formulation of quantum
dissipation/coherence, where the spectral density of the environment is related to 
the admittance of the isolation circuit. When the frequency of the qubit is at least
two times larger than the resonance frequency of the isolation circuit, we find
that the decoherence time of the qubit is two orders of magnitude larger than the
typical ohmic regime, where the frequency of the qubit is much smaller 
than the resonance frequency of the isolation circuit. 
Lastly, we show that when the qubit frequency is on resonance with the isolation circuit, 
an oscillatory non-Markovian decay emerges, as the dc SQUID phase qubit and its 
environment self-generate Rabi 
oscillations of characteristic time scales shorter than the decoherence time. 
\end{abstract}
\maketitle

The theoretical possibility of using quantum mechanics to manipulate information efficiently~\cite{feynman} 
has lead, through advances in technology, to the plausibility of building a quantum computer using
two-level systems, also called quantum bits or qubits. 
Several schemes have been proposed as attempts to manipulate qubits in atomic, molecular and optical physics (AMO),
and condensed matter physics (CMP). In AMO the most promising schemes are 
trapped ion systems~\cite{monroe-95}, and ultracold atoms in
optical lattices~\cite{brennen}.

On the CMP side, the pursuit of solid state qubits has been most promising in spin systems~\cite{hanson, hayashi}
and superconducting devices~\cite{devoret-02,lobb-01,shnirman-97}. While the manipulation
of qubits in AMO has relied on the existence of qubits in a lattice of ions or ultra-cold atoms and the use
of lasers, the manipulation of qubits in CMP has relied on the NMR techniques (spin qubits) and the Josephson
effect (superconducting qubits). Integrating qubits into a full quantum
computer requires a deeper understanding of decoherence effects in a single qubit and how different
qubits couple. 

In AMO systems Rabi oscillations in single qubits have been observed over time scales of miliseconds since each
qubit can be made quite isolated from its environment~\cite{monroe-95}, however it has been 
quite difficult to implement 
multi-qubit states as the coupling between different qubits is not yet fully controllable. 
On the other hand, in superconducting qubits Rabi oscillations have been observed~\cite{devoret-02} over 
shorter time scales (500ns), since these qubits are coupled to many environmental degreees of freedom, and
thus require very careful circuit design. Furthermore, extensions to the multi qubit regime also require 
further integrated circuit designs.

In this manuscript, we analyze decoherence effects in a single superconducting phase qubit 
coupled to isolation circuits. (Phase qubits~\cite{lobb-01} are superconducting qubits dominated by 
the Josephson effect, in contrast to charge qubits~\cite{shnirman-97}, which are dominated by quantization
of charge.)
The coupling of the qubit to the isolation circuit is emulated by a 
spectral density with an intrinsic resonance within the Caldeira-Leggett formulation of quantum dissipation.

\begin{figure}[htb]
\centerline{
\scalebox{0.42}{
\includegraphics{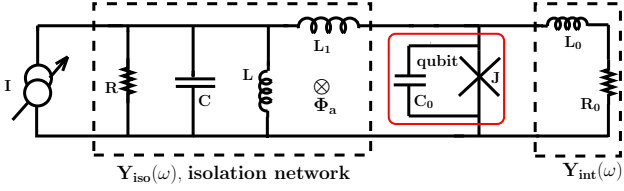}
}
}
\caption{Schematic drawing of the phase qubit with an RLC isolation circuit.}
\label{fig:one}
\end{figure}

The circuit used to describe intrinsic decoherence and self-induced Rabi oscillations
in phase qubits is shown in Fig.~\ref{fig:one}, which correponds to an asymmetric dc SQUID~\cite{martinis-02}.
The circuit elements inside the dashed box form an isolation network which serves two purposes:
a) it prevents current noise from reaching the qubit junction; 
b) it is used as a measurement tool.

The classical equation of motion for such a circuit is
\begin{equation}
\label{eqn:ceq1}
C_0 \ddot \gamma   + \frac{2\pi}{\Phi_0}I_{c0}\sin\gamma -\frac{2\pi}{\Phi_0}I + 
\int_0^t dt'  Y(t - t') \dot\gamma (t') = 0
\end{equation}
where $I_{c0}$ is the critical current of Josephson junction $J$ in Fig.\ref{fig:one}, and 
$\Phi_0 = h/2e$ is the flux quantum. The last term of Eq.~\ref{eqn:ceq1} can be written as
$i\omega Y(\omega) \gamma (\omega)$ in Fourier space.
The admittance function  $Y(\omega)$ can be modeled as two additive
terms $Y(\omega) = Y_{iso} (\omega) + Y_{int} (\omega)$. 
The first contribution $Y_{iso} (\omega)$ is the admittance that results when a transmission
line of characteristic impedance $R$ is attached to the isolation junction
(here represented by a capacitance $C$ and a Josephson inductance $L$)
and an isolation inductance $L_1$. Thus,
$
Y_{iso} (\omega) =  Z_{iso}^{-1} (\omega) 
$
where $Z_{iso} (\omega) = (i\omega L_1) + \left[  R^{-1} + i\omega C + (i\omega  L)^{-1} \right]^{-1}$
is the impedance of the isolation network shown in Fig.~\ref{fig:one}. 
The replacement of the isolation junction by an LC circuit is justified because under
standard operating conditions the external flux $\Phi_a$ varies to cancel the current flowing through 
the isolation junction making it zero biased~\cite{martinis-02}. 
Thus, the isolation junction behaves as a harmonic oscillator with
inductance $L$ which is chosen to be much smaller than $L_1$. 
The second contribution $Y_{int} (\omega)$ is an internal admittance 
representing the local environment of the qubit junction, such as defects in the oxide 
barrier, quasiparticle tunneling, or the substrate, and can be modeled by 
$Y_{int} (\omega) =  (R_0 + i\omega L_0)^{-1}$, where $R_0$ is the resistance
and $L_0$ is the inductance of the qubit as shown in Fig.~\ref{fig:one}.

Next, we use the Caldeira-Leggett formalism to describe the coupling of the phase qubit to the isolation network,
via the spin-boson Hamiltonian~\cite{leggett-87}
\begin{equation}
\label{eqn:hamiltonian}
\widetilde{H} = \frac{\hbar \omega_{01}}{2}\sigma_z
+ \sum_k \hbar \omega_k b_k^\dagger b _k
+ H_{SB},
\end{equation}
written in terms of Pauli matrices $\sigma_i$ (with $i = x, y, z$) and boson 
operators $b_{k}$ and $b_{k}^\dagger$.
The first term in Eq.~(\ref{eqn:hamiltonian}) represents a two-level approximation
for the phase qubit (system) described by states $\vert 0 \rangle$ and $\vert 1 \rangle$ 
with energy difference $\hbar \omega_{01} = \sqrt{8 E_c E_J} ( 1 - I^2/I_{c0}^2)^{1/4}$,
where $E_c = (2e)^2/2C$ is the capacitive energy, and $E_J = (\Phi_0/2\pi) I_{c0}$ is
the Josephson energy.
The second term corresponds to the isolation
network (bath) represented by a bath of bosons, where 
$b_{k}$ and $b_{k}^\dagger$ are the annihilation and creation operator of 
the $k$-th bath mode with frequency $\omega_{k}$. 
The third term is the system-bath (SB) Hamiltonian  
\begin{equation}
\label{eqn:system-bath-hamiltonian}
H_{SB} = 
\frac{1}{2}\sigma_x\hbar \langle 1 \vert \gamma \vert 0 \rangle
\sum_k \lambda_k \left( b_{k}^\dagger +  b_{k} \right).
\end{equation}
corresponding to the coupling between the isolation network and the
phase qubit which appears as $\int_0^t d t'S(t - t') \dot \gamma (t')$ in the classical
equation of motion Eq.~(\ref{eqn:ceq1}). 

The spectral density of the bath modes $J(\omega) = 
\hbar \sum_k \lambda_k^2  \delta \left( \omega- \omega_{k} \right)$ has dimensions 
of energy and can be written
as $J(\omega) = \omega {\rm Re} Y (\omega) (\Phi_0/2\pi)^2$, which leads to the compact form
$J(\omega) = J_{iso} (\omega) + J_{int} (\omega)$. The spectral
density of the isolation network is 
\begin{equation}
\label{eqn:spectral-density-isolation}
J_{iso}\left(\omega\right) = \left(\frac{\Phi_0}{2\pi} \right)^2
\frac{\alpha\omega}{\left(1-\omega^2/\Omega^2\right)^2+4\omega^2\Gamma^2/\Omega^4},
\end{equation}
where $\alpha=L^2/((L+L_1)^2 R) \approx (L/L_1)^2/R$ is the leading order term in the 
low frequency ohmic regime, $\Omega=\sqrt{(L+L_1)/(LL_1C)} \approx 1/\sqrt{LC}$ is essentially 
the resonance frequency, and $\Gamma=1/(2CR)$ plays the role of resonance width. 
Here, we used $L_1 \gg L$ corresponding to the relevant experimental regime.
Notice that $J_{iso} (\omega)$ has Ohmic behavior at low frequencies
$\lim_{\omega\rightarrow 0}J_{iso} (\omega)/\omega = (\Phi_0/2\pi)^2 (L/L_1)^2/R$, but 
has a peak at frequency $\Omega$ with broadening controlled by $\Gamma$.
In addition, notice that the dimensionless parameter 
$\Gamma/\Omega^2 = LL_1/\left(2R\left(L_1+L\right)\right) \approx R/L$
is independent of $C$. Therefore, when there is no capacitor ($C \to 0$), 
the resonance disappears and 
\begin{equation}
\label{eqn:sd-drude}
J_{iso} ( \omega ) =   \left(\frac{\Phi_0}{2\pi} \right)^2 \frac{\alpha\omega}{1+4\omega^2\Gamma^2/\Omega^4},
\end{equation}
reduces to a Drude term with characteristic frequency $\Omega^2/2\Gamma \approx  R/L$. 
The internal spectral density of the qubit is  
\begin{equation}
\label{eqn:spectral-density-internal}
J_{int}\left(\omega\right) = \left(\frac{\Phi_0}{2\pi} \right)^2
\frac {(\omega/R_0)} {1 + \omega^2 L_0^2/R_0^2}
\end{equation}
is a Drude term with characteristic frequency $R_0/L_0$.
Notice that $J_{int} (\omega)$ also has Ohmic behavior at low frequencies
$\lim_{\omega\rightarrow 0} J_{int} (\omega)/\omega = (\Phi_0/2\pi)^2/R_0$.
In order to obtain the relaxation $T_1$ and decoherence times $T_2$, 
we write the Bloch-Redfield equations
\begin{equation}
\label{eqn:d-matrix}
\dot{\rho}_{nm} = -i\omega_{nm}\rho_{nm} + \sum_{kl} R_{nmkl}\rho_{kl}
\end{equation}
for the density matrix $\rho_{nm}$ of the spin-boson Hamiltonian in 
Eq.~(\ref{eqn:hamiltonian}) and~(\ref{eqn:system-bath-hamiltonian})
derived in the Born-Markov limit.
Here all indices take the values $0$ and $1$ corresponding to the
ground and excited states of the qubit, respectively, while $\omega_{nm}=(E_n-E_m)/\hbar$
is the frequency difference between states $n$ and $m$.
The Redfield rate tensor is
\begin{equation}
R_{nmkl} = -\Gamma_{lmnk}^{(1)}-\Gamma_{lmnk}^{(2)} + \delta_{nk} \Gamma_{lrrm}^{(1)}  
 + \delta_{lm} \Gamma_{nrrk}^{(2)},  
\label{eq:redfield-tensor}
\end{equation}
where repeated indices indicate summation, and 
\begin{equation}
\Gamma_{lmnk}^{(j)} = 
\hbar^2  \int_0^\infty dt   
e^{-i\eta_j t} \langle H_{SB,lm}(t) H_{SB,nk}(0)\rangle, 
\end{equation}
where $j =1, 2$ and $\eta_1 = \omega_{nk}$ amd $\eta_2 = \omega_{lm}$.
Under these conditions, the relaxation rate $1/T_1 = \sum_n\limits R_{nnnn}$
becomes
\begin{equation}
\label{eqn:t1}
\frac{1}{T_1} = \frac{1}{M\omega_{01}} J(\omega_{01}) \coth \left( \frac{\hbar\omega_{01}}{k_BT} \right), 
\end{equation}
where $M \equiv \left(\Phi_0/2\pi\right)^2C_0$ has dimensions of mass $\times$ area (or energy $\times$ time squared) 
and is refered to as the {\it mass} of the phase qubit with capacitance $C_0$, while $\omega_{01}$ is the 
qubit frequency.
The interpretation of $T_1^{-1}$ is as follows: for the system to make a transition it needs to exchange 
energy $E = \hbar \omega_{01}$ with the environment using a single boson. 
The factor $ \coth (\hbar \omega_{01}/k_B T) =   n(\omega_{01})+ 1 + n(\omega_{01})$ captures 
the sum of the rates for emission 
(proportional to $n(\omega_{01})+ 1$) and absorption (proportional to $n(\omega_{01})$ of 
a boson), where $n(\omega_{01}) = \left[ \exp (\hbar \omega_{01}/k_B T) - 1 \right]^{-1}$ is the Bose function.

\begin{figure}
\centerline{ \scalebox{0.7} {\includegraphics{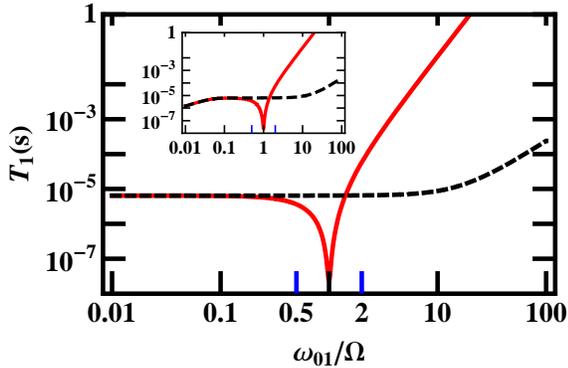}} } 
\caption{\label{fig:two}
$T_1$ (in seconds) as a function of qubit frequency 
$\omega_{01}$. The solid (red) curves describes an RLC isolation network with
parameters $R = 50$~ohms, $L_1=3.9$nH, $L=2.25$pH, $C=2.22$pF, and 
qubit parameters $C_0=4.44$pF,  $R_0 = \infty$ and $L_0 = 0$. 
The dashed curves correspond to an RL isolation network with the same parameters,
except that $C = 0$. Main figure ($T =0$), inset ($T = 50$mK) with $\Omega = 141$ GHz.
}
\end{figure}

In Fig.~\ref{fig:two}, $T_1$ is plotted versus qubit frequency $\omega_{01}$ 
for spectral densities describing an RLC (Eq.~\ref{eqn:spectral-density-isolation}) 
or Drude (Eq.~\ref{eqn:sd-drude}) isolation network at fixed temperatures $T = 0$ (main figure) 
and $T = 50$mK (inset), for $J_{int} (\omega) = 0$ corresponding to $R_0 \to \infty$.
In the limit of low temperatures $(k_B T/\hbar \omega_{01} \ll 1)$, the
relaxation time becomes $T_1 (\omega_{01}) = M \omega_{01}/J(\omega_{01})$. 
From Fig.~\ref{fig:two} (main plot) several important points can be extracted. 
First, in the low frequency regime ($\omega_{01} \ll \Omega)$ the RL (Drude) and RLC environments 
produce essentially the same relaxation time $T_{1,RLC} (0) = T_{1,RL} (0) = T_{1,0} 
\approx (L_1/L)^2 R C_0$, because both systems are ohmic. 
Second, near resonance ($\omega_{01} \approx \Omega$), $T_{1,RLC}$ is substantially reduced 
because the qubit is resonantly coupled to its environment producing a distinct non-ohmic behavior.
Third, for ($\omega_{01} > \Omega$), $T_1$ grows very rapidly in the RLC case. 
Notice that for $\omega_{01} > \sqrt{2} \Omega$, the RLC relaxation time $T_{1,RLC}$ is always larger than $T_{1,RL}$.
Furthermore, in the limit of $\omega_{01} \gg {\rm max} \lbrace {\Omega, 2\Gamma} \rbrace$, 
$T_{1,RLC}$ grows with the fourth power of $\omega_{01}$ behaving as $T_{1,RLC} \approx T_{1,0} \omega_{01}^4/\Omega^4$,
while for $\omega_{01} \gg \Omega^2/2\Gamma$, $T_{1,RL}$ grows only with second power of $\omega_{01}$ behaving
as $T_{1,RL} \approx 4 T_{1,0} \Gamma^2 \omega_{01}^2/\Omega^4$. Thus, $T_{1,RLC}$ is always much larger than
$T_{1,RL}$ for sufficiently large $\omega_{01}$. Notice, however, that for parameters in the experimental range
such as those used in Fig~\ref{fig:two}, $T_{1,RLC}$ is two orders of magnitude larger than $T_{1,RL}$, 
indicating a clear advantage of the RLC environment shown in Fig~\ref{fig:one} over the standard ohmic RL environment.
Thermal effects are illustrated in the inset of Fig.~\ref{fig:two} where $T = 50$mK is a characteristic temperature
where experiments are performed~\cite{paik-07}. The typical values of $T_1$ at low frequencies vary from 
$10^{-5}$s at $T =0$ to $10^{-6}$s at $T = 50$mK, while the high frequency values remain
essentially unchanged as the thermal effects are not important for $\hbar \omega_{01} \gg k_B T$. 

\begin{figure}
\centerline{ \scalebox{0.65} {\includegraphics{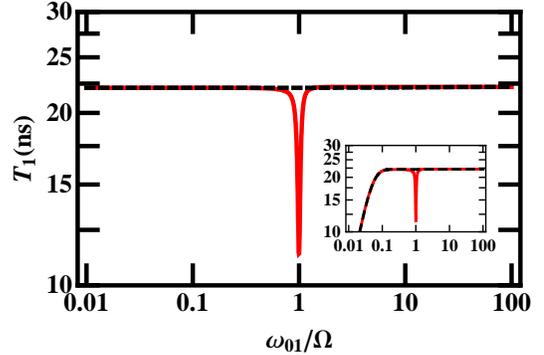}} } 
\caption{\label{fig:three}
$T_1$ (in nanoseconds) as a function of qubit frequency 
$\omega_{01}$.
The solid (red) curves describes an RLC isolation network with
parameters $R = 50$~ohms, $L_1=3.9$nH, $L=2.25$pH, $C=2.22$pF, and 
qubit parameters $C_0=4.44$pF, $R_0 = 5000$~ohms and $L_0 = 0$. 
The dashed curves correspond to an RL isolation network with the same parameters,
except that $C = 0$. Main figure ($T =0$), inset ($T = 50$mK) with $\Omega = 141$ GHz.}
\end{figure}

In the preceeding analysis we neglected the effect of the local environment by 
setting $Y_{int} (\omega) = 0$. As a result, the low-frequency value of $T_1$ is
substantially larger than obtained in experiment~\cite{paik-07, martinis-02}. By modeling
the local environment with $R_0 = 5000$~ohms and $L_0 = 0$ we obtain the $T_1$ 
versus $\omega_{01}$ plot shown in Fig.~\ref{fig:three}.
Notice that this value of $R_0$ brings $T_1$ to values close to $20$ns at $T = 0$.
The message to extract from Figs.~\ref{fig:two} and~\ref{fig:three} is that increasing $R_0$ as much as 
possible and increasing the qubit frequency $\omega_{01}$ from $0.1\Omega$ to $2\Omega$ at fixed
low temperature can produce a large increase in $T_1$.

Although the Bloch-Redfield equations described in Eq.~(\ref{eqn:d-matrix})
capture the long time behavior of the density matrix, they can not describe the short time behavior
of the system in particular near resonance where $\omega_{01} \approx \Omega$, where
the RLC spectral density is very large. In this case, only the environmental
modes with $\omega_{k} = \Omega$ couple strongly to the two-level system, like a two-level atom coupled to
an electromagnetic field cavity mode that has a finite lifetime.
This is best seen by restricting the Hamiltonian described in Eqs.~(\ref{eqn:hamiltonian}) 
and~(\ref{eqn:system-bath-hamiltonian}) only to boson modes with $\omega_{k} \approx \Omega \approx \omega_{01}$.
In this case, it is best to rewrite the spectral density as 
\begin{equation}
\label{eqn:sd-poles}
J_{iso} (\omega) = \left( \frac{\Phi_0}{2\pi} \right)^2 
\frac{\alpha\Omega^3}{4\text{i}\Gamma}
\sum_{\sigma=\pm 1}\frac{\sigma\omega}{\omega^2-\left(\sigma\widetilde{\Omega} + \text{i}\Gamma\right)^2},
\end{equation}
where $\alpha$ has the same definition as in Eq.~(\ref{eqn:spectral-density-isolation}),
and $\widetilde\Omega = \Omega -\Gamma^2/\Omega$.
This reveals a resonance at $\omega = \widetilde \Omega$ with linewidth $\Gamma$, such
that $J_{int} (\omega = \widetilde \Omega)$ can be neglected for any non-zero 
value of $R_0$, and $J (\omega) \approx J_{iso} (\omega)$. 

When $\omega_{k} \approx \Omega \approx \omega_{01}$, the Hamiltonian in  
Eqs.~(\ref{eqn:hamiltonian}) and~(\ref{eqn:system-bath-hamiltonian}) can be solved
in the rotating wave approximation using the complete basis set of system-bath product states 
$\vert \psi_0 \rangle = \vert 0 \rangle_{\rm S} \otimes \vert 0\rangle_{\rm B}$;
$\vert \psi_1 \rangle = \vert 1 \rangle_{\rm S} \otimes \vert 0\rangle_{\rm B}$; 
$\vert \psi_k \rangle = \vert 0\rangle_{\rm S} \otimes \vert k \rangle_{\rm B}$, where
$\vert 0 \rangle_{\rm S}$ and $\vert 1 \rangle_{\rm S}$ are the states of the
qubit and $\vert k \rangle_{\rm B}$ are the states of the bath.
Hence, the state of
the total system at any time is 
\begin{equation}
\label{eq:phi_exp}
\phi(t)=c_0\psi_0+c_1(t)\psi_1+\sum_k c_k(t)\psi_k,
\end{equation}
with probability amplitudes $c_0$, $c_1(t)$, and $c_k(t)$. 
The amplitude $c_0$ is constant, while the amplitudes $c_1(t)$ and $c_k(t)$ are time dependent. 
Assuming that there are no excited bath modes at $t = 0$, we impose the
initial condition $c_k(0) = 0$, and use the normalization 
$\vert \phi (t) \vert^2 = 1$ to obtain the closed integro-differential equation
\begin{equation}
\dot{c}_1 (t) = - \int_0^t dt_1 f(t-t_1) c_1(t_1),
\end{equation}
where the kernel is the correlation function
$$
f(\tau) = \int d\omega J(\omega) \left[ \coth\left(\frac{\hbar \omega}{2k_B T}\right)\cos(\omega \tau)
+ i \sin(\omega \tau) \right]
\nonumber
$$
directly related to the spectral density $J(\omega)$.
In the present case the reduced density matrix is
\begin{equation}
\label{eq:rho_ex}
\rho(t)=
\left(
\begin{array}{cc}
|c_1(t)|^2  & c_1(t) c_0^*\\
c_1^*(t)c_0 & |c_0|^2+\sum_k|c_k(t)|^2
\end{array}
\right) .
\end{equation}
which in combination with the condition that
$ |c_0|^2+\sum_k|c_k(t)|^2 = 1  -  |c_1(t)|^2 $ (${\rm Tr} \rho (t) = 1$)
indicates that the time dynamics of $\rho(t)$ is fully determined
by $c_1 (t)$.

\begin{figure}[htb]
\centerline{ \scalebox{0.7} {\includegraphics{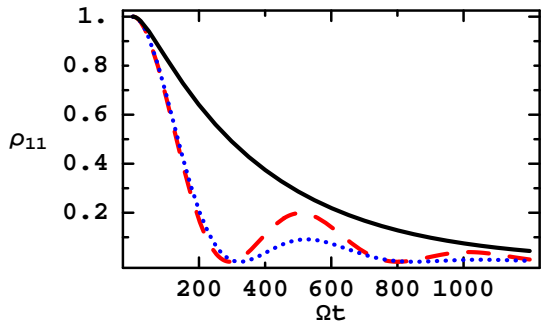}} } 
\caption{\label{fig:four}
Population of the excited state of the qubit as a function of time 
$\rho_{11}(t)$, with $\rho_{11} (t = 0) = 1$ for $R = 50$~ohms (solid curve),
$350$~ohms (dotted curve), and $R = 550$~ohms (dashed curve), and
$L_1=3.9$nH, $L=2.25$pH, $C=2.22$pF, $C_0=4.44$pF, $R_0 = \infty$ and $L_0 = 0$.}
\end{figure}

In the limit of $T = 0$, we can solve for $c_1(t)$ exactly and
obtain the closed form
$$
c_1(t)=
{\cal L}^{-1} 
\left\{
\frac{ (s+\Gamma - i\omega_{01} )^2 + \Omega^2 - \Gamma^2}
{s \left[ (s + \Gamma - i\omega_{01} )^2 + \Omega^2 - \Gamma^2 \right] - \kappa \Omega^4\pi i/\Gamma}
\right\}
$$
where ${\cal L}^{-1} \lbrace F (s) \rbrace$ is the inverse Laplace transform
of $F(s)$, and $\kappa = (\alpha/M \omega_{01}) \times (\Phi_0/2\pi)^2 \approx 1/(\omega_{01} T_{1,0})$.
The element $\rho_{11} = \vert c_1 (t) \vert^2$ of the density matrix
is plotted in Fig.~\ref{fig:four} for three different values of resistance,
assuming that the qubit is in its excited state such that $\rho_{11} (0) = 1$.
We consider the experimentally relevant limit of $\Gamma \ll \omega_{01} \approx \Omega$,
which corresponds to the weak dissipation limit. Since $\Gamma = 1/(2CR)$ the width 
of the resonance in the spectral density shown in Eq.~(\ref{eqn:sd-poles}) is 
smaller for larger values of $R$. Thus, for large $R$, the RLC environment 
transfers energy resonantly back and forth to the qubit and induces
Rabi-oscillations with an effective time dependent decay rate 
$
\gamma(t) = - 2\Re \left \{ \dot c_1(t) / c_1(t) \right\}.
$

These environmentally-induced Rabi oscillations are a clear signature of the non-Markovian 
behavior produced by the RLC environment, and are completely absent in the RL environment
because the energy from the qubits is quickly dissipated without being temporarily stored.
These environmentally-induced Rabi oscillations are generic features of 
circuits with resonances in the real part of the admittance. 
The frequency of the Rabi oscillations $\Omega_{Ra} = \sqrt{\pi \kappa \Omega^3/2\Gamma}$
is independent of the resistance since $\Omega_{Ra} \approx \Omega \sqrt{\pi L^2 C/L_1^2 C_0}$,
and has the value of $\Omega_{Ra} = 2\pi f_{Ra} \approx 360 \times 10^{6}$~rad/sec for Fig.~\ref{fig:four}.

In conclusion, we analyzed decoherence effects in a single superconducting phase qubit 
coupled to isolation circuits with an intrinsic resonance,
and emphasized the crucial role played by the design of the isolation circuit on
decoherence properties. In particular, for an RLC isolation cicuit, we found 
that the decoherence time of the qubit is 
two orders of magnitude larger than its typical low frequency ohmic regime, 
provided that the frequency of the qubit is about two times larger than the resonance frequency 
of the isolation circuit. 

Lastly, we showed that when
the qubit frequency is close to resonance with the isolation circuit, the non-oscillatory
Markovian decay of the excited state population of the qubit, gives in to an oscillatory
non-Markovian decay, as the phase qubit and its environment self-generate Rabi 
oscillations of characteristic time scales shorter than the decoherence time.
 \acknowledgements{We acknowledge support from NSF (DMR-0304380) and NSA, 
through the Laboratory of Physical Sciences.}

\end{document}